\documentclass[pra,twocolumn,aps,groupedaddress,showpacs]{revtex4}
\usepackage{amsmath,amsfonts,amssymb}
\usepackage{wrapfig}
\usepackage{graphicx}
\usepackage{bm,epsfig} % bbm excluded; use \opeone 
\usepackage{graphics}
\newcommand{\be}{\begin{eqnarray}}
\newcommand{\ee}{\end{eqnarray}}

\newcommand{\bra}[1]{\langle#1|}
\newcommand{\ket}[1]{|#1\rangle}

\newcommand{\affUni}{Institut f\"ur Theoretische Physik, Universit\"at Innsbruck, Technikerstr.~25, 6020~Innsbruck, Austria}
\newcommand{\affIQOQI}{Institut f\"ur Quantenoptik und Quanteninformation der \"Osterreichischen Akademie der Wissenschaften, Innsbruck, Austria}

\begin{document}

%\bibliographystyle{apsrev}

%%%%%%%%%%%%%%%%%%%%%%%%%%%%%%%%%%%%%%%%%%%%%%%%%%%%%%%%%%%%%%%%%%%%%%%%%%%%%%

\title{Entanglement purification protocols for all graph states}

\author{Caroline Kruszynska}
\email{caroline.kruszynski@uibk.ac.at}
\affiliation{\affUni}
\affiliation{\affIQOQI}

\author{Akimasa Miyake}
\affiliation{\affUni}
\affiliation{\affIQOQI}

\author{Hans J. Briegel}
\affiliation{\affUni}
\affiliation{\affIQOQI}

\author{Wolfgang D\"ur}
\affiliation{\affUni}
\affiliation{\affIQOQI}

\date{June 9, 2006}

\begin{abstract}
We present multiparty entanglement purification protocols that are capable of purifying arbitrary graph states directly. We develop recurrence and breeding protocols and compare our methods with strategies based on bipartite entanglement purification in static and communication scenarios. We find that direct multiparty purification is of advantage with respect to achievable yields and minimal required fidelity in static scenarios, and with respect to obtainable fidelity in the case of noisy operations in both scenarios.
\end{abstract}


\maketitle

\section{Introduction}

Entanglement purification (distillation) is an important primitive in quantum information processing. It allows one to overcome the influence of noise in quantum communication \cite{Be96,BDSW96,De96}, and enables one to obtain provable secure quantum key distribution even in the context of noisy channels \cite{As00} and over arbitrary distances \cite{Br98}. The applications of entanglement purification are however not limited to bipartite communication scenarios. First, it has been extended to certain multiparty scenarios \cite{Mu03,MaSm02,DAB03,ADB05,Ch04}, leading e.g. to novel quantum primitives such as multiparty secure state distribution \cite{Du05}. Second, in the context of fault-tolerant quantum computation, entanglement purification is a key ingredient to obtain improved error thresholds \cite{Du02QC,Kn05}. Applications in quantum error correction \cite{He06} and quantum simulation \cite{Du06} have also been discussed.

The basic idea of entanglement purification is to use several copies of a noisy entangled state to generate, by means of local operations and classical communication (LOCC), a few copies with improved fidelity. So far, entanglement purification protocols have been developed that are capable of purifying Bell states \cite{Be96,BDSW96,De96} in the bipartite case, and all two-colorable graph states \cite{DAB03,ADB05} as well as W states \cite{MB05} in the multiparty case. They also have been generalized to higher dimensional systems \cite{DN03}. Graph states \cite{HEB04,He06} are a family of multiparty entangled states with interesting entanglement properties. Two-colorable graph states are sub-family associated with two-colorable graphs and include a number of interesting states such as GHZ states, cluster states and codewords of CSS error correction codes. Graph states appear for instance in the context of measurement-based quantum computation, where a given graph state represents an algorithmic-specific resource that allows one to realize a specific unitary operation on several qubits by local measurements. Typically, these graphs are not two-colorable, but one may wish to purify this resource, e.g. to realize one-way quantum computation in a fault-tolerant manner. 

In this paper, we present entanglement purification protocols (EPP) that are capable of purifying {\em arbitrary} graph states. To be precise, we develop for each graph state a direct multiparty recurrence protocol and a multiparty breeding protocol. 
Our key ideas are twofold. First, we show that if auxiliary (even noisy) 
{\em two-colorable} graph states are available, 
the fidelity of the noisy graph states can be improved.
Second, we show how to get a single copy of the auxiliary two-colorable graph state from two identical copies of the $k$-colorable graph state by LOCC. 
Thus, unlike the known two-colorable graph state entanglement purification protocol \cite{DAB03,ADB05}, 
our protocol utilizes different shapes of graphs.

The protocols are applicable both in (i) a static (LOCC) scenario, where the 
parties attempt to purify by LOCC several copies of given noisy multiparty 
entangled states, as well as in (ii) a communication scenario where the 
parties are allowed to generate (arbitrary) multiparty states locally, distribute them through noisy quantum channels and attempt to 
end up with high-fidelity target graph states.
Our first idea is commonly utilized to both scenarios. The second idea is 
crucial in the static (LOCC) scenario, while in the communication scenario we 
may prepare separately auxiliary two-colorable graph states through noisy 
channels in an effective way.

In both cases, we show that the new direct multiparty EPP are superior to alternative approaches based on bipartite entanglement purification. In particular, we find that in the static scenario (i) the yield of direct multiparty breeding protocols is higher than for any strategy based on bipartite purification, and also the purification regime is larger. When also considering noisy local control operations, we show that for both scenarios (i) and (ii), the new multiparty recurrence protocol allows one to reach higher fidelities.
We first review the concept of graph states in Sec.~\ref{graph states:basics}, and then present new recurrence and breeding protocols in Sec.~\ref{sec:recurrence protocol} and \ref{sec:hashing}. We present an alternative strategy based on purification of two-colorable sub graph-states in Sec.~\ref{sec:alternative}. A comparison with bipartite strategies for noiseless in Sec.~\ref{sec: comparison with BEPP, perfect LO} and noisy local control operations in Sec.~\ref{sec: comparison with BEPP, imperfect LO} finally demonstrates the advantage of these new protocols.

%-------------------------------------------------------------------------------------------------
\section{Graph states and Manipulation}
%-------------------------------------------------------------------------------------------------
In this Section, we summarize the basics concerning graph states and their manipulations.

%-------------------------------------------------------------------------------------------------
\subsection{Definition and notation for graph states} \label{graph states:basics}
%-------------------------------------------------------------------------------------------------

Graph states are a family of multiparty entangled states associated with mathematical graphs \cite{He06}. 
A graph $G=(V,E)$ is given by a set $V=\{1,2,\ldots ,N\}$ of $N$ vertices connected in a specific way by edges $E$.
To every such graph there corresponds a basis of $N$--qubit states $\{|{\bm \mu}\rangle_G\}$, where each of the basis states $|{\bm \mu}\rangle_G$ $({\bm \mu} = \mu_1\mu_2\ldots \mu_N)$ is the common eigenstate of $N$ commuting correlation operators $K_a^G$ with eigenvalues  $(-1)^{\mu_a}$ such that $\mu_a = 0,1$. That is, they fulfill the set of eigenvalue equations $K_a^G |{\bm \mu}\rangle_G = (-1)^{\mu_a}|{\bm \mu}\rangle_G$, $a=1,\ldots,N$. The correlation operators are uniquely determined by the graph $G$ and are given by 
\be
\label{eq:K}
K_a^G= \sigma_x^{(a)} \prod_{\{a,b\} \in E} \sigma_z^{(b)},
\ee
where $\sigma_\alpha^{(a)}$ denotes the application of the corresponding Pauli operator $(\alpha=x,y,z)$ by the party $a$. Equivalently, any graph state $\ket{\bm{\mu}}_G$ can be written in the following manner:
\be \label{eq:graph states: interaction picture}
\ket{\bm{\mu}}_G=\prod_{a=1}^N(\sigma^{(a)}_z)^{\mu_a}\left(\prod_{\{b,c\}\in E}\Lambda Z^{(bc)}\right)\ket{+}^{\otimes N}
\ee
where $\Lambda Z=\operatorname{diag}(1,1,1,-1)$ in the computational basis 
is the controlled-phase gate, and 
$\ket{\pm} = \frac{1}{\sqrt{2}}(\ket{0}\pm\ket{1})$. 

We will use the concept of $k$-coloration in the following. A graph is called $k$--colorable if there exist $k$ sets of vertices $A_1,A_2,\ldots, A_k\in V$ such that there are no edges within each of the groups $A_j$ for all $j$, i.e. for all $a,b \in A_j$, and for all $j$, we have $\{a,b\} \not\in E$. Two-colorable graphs are a special instance with $k=2$, where multiparty entanglement purification protocols are known \cite{DAB03,ADB05}. However, only a subset of graphs is two-colorable and, in principle, a graph may be $N$-colorable. We remark that local-unitary equivalent graph states \cite{HEB04} may correspond to graphs with different coloring, and the minimum $k$ within the local equivalence class is not known. Under local Clifford operation, we have however that generally $k >2$, i.e. not all graph states are locally equivalent to two-colorable graphs.

\begin{figure}
\setlength{\unitlength}{0.05625\columnwidth}
\begin{picture}(16,15)
\thinlines
\put(0,1.5){\includegraphics[width=0.9\columnwidth]{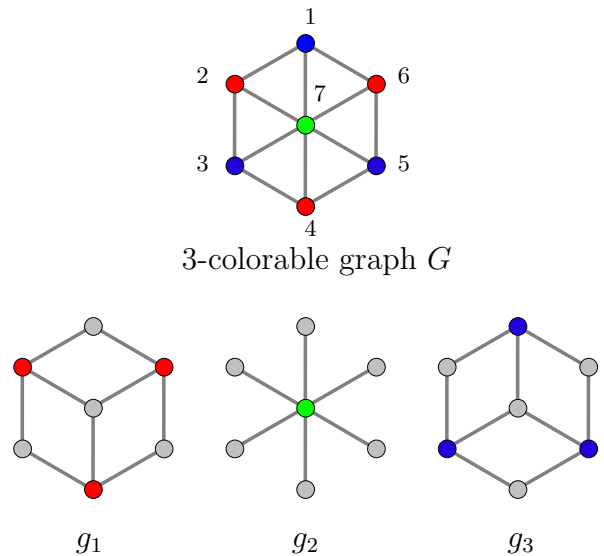}}
\put(5,10.5){3}
\put(5,12.9){2}
\put(7.95,14.5){1}
\put(10.5,12.9){6}
\put(10.5,10.5){5}
\put(7.95,8.7){4}
\put(8.2,12.4){7}
\put(4.6,7.8){\mbox{\large 3-colorable graph $G$}}
\put(1.7,0.2){\mbox{\large $g_1$}}
\put(7.6,0.2){\mbox{\large $g_2$}}
\put(13.5,0.2){\mbox{\large $g_3$}}
\end{picture}
\caption{(Color online). A 3-colorable graph $G$ and the 3 corresponding two-colorable sub-graphs $g_1$, $g_2$, and $g_3$. $g_1$ corresponds to the red color (vertices 2,4 and 6), $g_2$ to the green color (vertex 7) and $g_3$ to the blue (vertices 1,3 and 5).
}
\label{fig:G and gj}
\end{figure}

Associated with a given $k$-colorable graph $G$ with coloring $\{A_1,A_2,\dots,A_k\}$, we define two-colorable graphs $\{g_1,g_2,\dots,g_k\}$ (see Fig.\ref{fig:G and gj} for an illustration), where $g_j$ contains only the edges between the set $A_j$ and the remaining sets $\{A_i, i\not =j\}$, but where edges between the remaining sets are erased. That is, the sets $\{A_i, i\not=j \}$ form a new set $A_{\bar j}=V\setminus A_j$. The set of indices $\mu_a$ corresponding to the set $A_{\bar j}$ will be denoted by ${\bm \mu_{\bar j}}$. Note that $\cup_{j=1}^k g_j=G$.

In this paper, we consider mixed states diagonal in the graph-state basis,
\be
\label{mixedgraph}
\rho_G = \sum_{\bm \mu}\lambda_{\bm \mu_1\bm \mu_2 \ldots \bm\mu_k} |{\bm \mu_1\bm \mu_2\ldots \bm\mu_k}\rangle_G\langle {\bm \mu_1\bm\mu_2\ldots \bm\mu_k}|,
\ee
where we have grouped the multi-index ${\bm \mu}$ into $k$ multi-indices ${\bm \mu_j}=\mu_{{j_1}}\ldots \mu_{{j_m}}$ ($m=|A_j|$) corresponding to the sets $A_j$ defined by a chosen $k$-coloration of the graph $G$, note that mixed states resulting from any noise models can be brought to this form by means 
of local depolarization, i.e. by applying randomly the local operations 
corresponding to the correlation operators $\{K^G_a\}$. 
Diagonal elements in the graph-state basis, i.e. the coefficients $\lambda_{\bm \mu_1\bm \mu_2 \ldots \bm\mu_k}$ remain unchanged by this procedure, and
any mixed state can be assumed to be of the form (\ref{mixedgraph}) without 
loss of generality.
Hence we can interpret the mixed state as an ensemble of graph states $\ket{\bm{\mu}}_G$, where $ \ket{\bm{\mu}}_G$ appears  with probability $\lambda_{\bm \mu}$. We can therefore restrict our attention to the transfer of these indices 
between unknown pure states as presented in the next subsection.

%-------------------------------------------------------------------------------------------------
\subsection{Operations on graph states}\label{Seq:operations on graph states}
%-------------------------------------------------------------------------------------------------
We briefly mention two operations which play key roles in entanglement 
purification protocols. One is the multilateral CNOT which enables one to 
transfer the stabilizer eigenvalues ${\bm \mu}$ between two states.
The other are Pauli measurements which allow one to evaluate ${\bm \mu}$ 
with the help of classical communication.

\begin{figure}
\setlength{\unitlength}{0.05625\columnwidth}
\begin{picture}(16,5)
\thinlines
\put(0,1.5){\includegraphics[width=0.9\columnwidth]{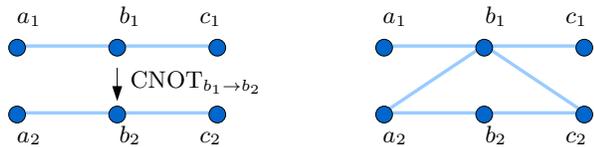}}
\put(0.0,4.25){\mbox{ $a_1$}}
\put(2.8,4.25){\mbox{ $b_1$}}
\put(5,4.25){\mbox{ $c_1$}}
\put(0.0,1.){\mbox{ $a_2$}}
\put(2.8,1.){\mbox{ $b_2$}}
\put(5,1.){\mbox{ $c_2$}}
\put(10,4.25){\mbox{ $a_1$}}
\put(12.8,4.25){\mbox{ $b_1$}}
\put(15,4.25){\mbox{ $c_1$}}
\put(10.0,1.){\mbox{ $a_2$}}
\put(12.8,1.){\mbox{ $b_2$}}
\put(15,1.){\mbox{ $c_2$}}
\put(3.1,2.5){\mbox{ $\operatorname{CNOT}_{b_1\rightarrow b_2}$}}
\end{picture}
\caption{(Color online). Effect of a CNOT gate on a product of two graph-states. The final state is associated to a graph composed of the two initial graphs with additional edges between the control qubit $b_1$ and all neighbors of the target qubit $b_2$.}
\label{fig:cnot produces edges}
\end{figure}

Let us first describe the action of a CNOT gate on a product of two (possibly different) $N$-qubit graph-states $\ket{\bm{\mu}}_{G_1}$ and $\ket{\bm{\nu}}_{G_2}$. We apply the CNOT gate from $a_1$ ($a^{th}$ qubit of the first state) to $a_2$ ($a^{th}$ qubit of the second state). We deduce the expression for the resulting state from (\ref{eq:graph states: interaction picture}). As $\ket{+}\otimes\ket{+}$ is an eigenvector of CNOT associated to the eigenvalue 1, the difference between the initial and the final state is due to the commutation relation between the CNOT gate, the $\sigma_z$ operators and the controlled phase gates. A straightforward calculation yields

\begin{align}
\label{eq:CNOT}
& \operatorname{CNOT}^{(a_1\rightarrow a_2)}\ket{\bm{\mu}}_{G_1}\ket{\bm{\nu}}_{G_2}= 
\nonumber\\
&\qquad 
(\sigma_z^{(a_1)})^{\nu_{a_2}}\prod_{a'_2\in N_{a_2}}\Lambda Z^{(a_1 a'_2)}\ket{\bm{\mu}}_{G_1}\ket{\bm{\nu}}_{G_2},
\end{align}
where $N_{a_2}$ are the neighbors of vertex $a_2$. The final state is related to a graph of $2\,N$ vertices which is composed of the two initial graphs with addition of all edges between vertex $a_1$ and the neighbors of vertex $a_2$. See Fig~\ref{fig:cnot produces edges} for an illustration. In addition, bit $\mu_{a_1}$ is flipped if bit $\nu_{a_2}=1$, which corresponds to a transfer of information from the second state to the first one. Note however that this parity information cannot be evaluated without disturbing the first state. 

To read out the information transfered between two states, one uses 
Pauli measurements in directions that correspond to the correlation 
operators $\{K^G_a\}$. 
Suppose we want to determine the bit string $\bm{\mu}_j$ corresponding to 
a color $A_j$ for a given graph state 
$\ket{\bm{\mu}_j,\bm{\mu}_{\bar{j}}}_G$.  
Then, the parties belonging to color $A_j$ measure their qubit $a$ in the eigenbasis $\{\ket{\pm}=\frac{1}{\sqrt{2}}(\ket{0}\pm \ket{1})\}$ of $\sigma_x$, obtaining results $\xi_a\in\{0,1\}$, while all the other parties make their measurement in the eigenbasis $\{\ket{0},\ket{1}\}$ of $\sigma_z$, obtaining results $\zeta_b\in\{0,1\}$.
According to Eq.~(\ref{eq:K}), each stabilizer eigenvalue $\mu_a$ of 
$K^G_a$ $(a\in A_j)$ can be evaluated via classical communication of these 
measurement results as,
\be \label{eq:measurement}
\mu_a =  \xi_a \bigoplus_{b\in N_a} \zeta_b.
\ee
Note that all indices corresponding to color $A_j$ are measured simultaneously.

%-------------------------------------------------------------------------------------------------
\section{Static scenario and communication scenario}
%-------------------------------------------------------------------------------------------------
In this article, we are interested in the generation of high-fidelity graph states shared among $N$ spatially separated parties. We consider two cases (i) a static LOCC scenario, and (ii) a communication scenario,
since both situations are of practical relevance in quantum information processing.

In (i), the static LOCC scenario, the parties share $M$ copies of a mixed state $\rho$. They can manipulate the states by means of LOCC in order to generate a high fidelity approximation of the target state $\ket{\bm 0}_G$. No quantum communication between the parties is allowed. The efficiency of this procedure is measured by the yield, i.e. the ratio of high fidelity output state per input state. 
In (ii), the communication scenario, the parties are connected by noisy quantum channels and are allowed to distribute any locally generated entangled state through the noisy quantum channels, in addition to sequences of LOCC.
The efficiency of this procedure is measured by the (inverse of the) quantum communication cost \cite{KADB05}, i.e. the number of channel usages required to generate a desired multiparty entangled state with sufficiently high fidelity.

The scenario (i) deals with entanglement properties of a given mixed state $\rho$, i.e. whether high fidelity multiparty entangled states can be distilled from several copies of $\rho$ by LOCC. The scenario (ii), on the other hand, has a certain physical set-up (a communication set-up) in mind, and hence deals with the question whether in such a context high-fidelity entangled states can be generated. For our purposes, the most relevant difference between the two scenarios is that in case of (ii) noisy entangled states of {\em any} kind can be generated by distributing them through noisy quantum channels, which can be used to generate the desired target state. In (i), only copies of $\rho$ are given which should be manipulated by LOCC. If different states are needed, this has to be prepared by manipulating several copies of $\rho$ by LOCC.

%-------------------------------------------------------------------------------------------------
\section{Direct Multipartite Entanglement Purification Protocol}
%-------------------------------------------------------------------------------------------------
%-------------------------------------------------------------------------------------------------
\subsection{Recurrence protocol} \label{sec:recurrence protocol}
%-------------------------------------------------------------------------------------------------
We now present a recurrence protocol that allows one to purify directly any $k$--colorable graph-state, provided auxiliary two-colorable graph-states corresponding to the $k$ different colors are available. The sub-protocol is applicable in both scenarios, and we describe later on how to construct the auxiliary two-colorable graph-states in each scenario. Note that if the target graph state is two-colorable, auxiliary graphs $g_1$ and $g_2$ are nothing but $G$. Thus our protocol covers known protocols for two-colorable graph states.  The total protocol consists of $k$ sub-protocols $\{\mathcal{P}_j\}  \;(j=1,\dots,k)$, each of which serves to purify a state partially, i.e. with respect to the index vector ${\bm \mu_j}$ corresponding to a group $A_j$ (i.e. a certain color). We outline the protocol in the following.

\subsubsection{Sub-protocol $\mathcal{P}_j$}

Assume that auxiliary (possibly noisy) two-colorable graph-states $\rho_{g_j}$ 
are available in addition to an ensemble of $k$-colorable graph-states
 $\rho_G$. We purify the $k$-colorable graph-states $\rho_G$ with respect 
to bit string ${\bm \mu}_j$:

\begin{itemize}
\item[(1-1)] The parties take two noisy states in such a way that the first state
is $\rho_G$ and the second one is $\rho_{g_j}$. The parties in the group $A_j$ 
apply CNOT gates from the second (control) state $\rho_{g_j}$ to the first 
(target) state $\rho_G$, 
while the parties in the group $A_{\bar{j}}$ apply the CNOT gate in the 
opposite direction.

\item[(1-2)] The parties measure the second state $\rho_{g_j}$ locally in the eigenbasis of $\sigma_x$ for the group $A_j$ and in the eigenbasis of $\sigma_z$ for the group 
$A_{\bar{j}}$.
Using classical communication, they decide to keep the first state $\rho_G$ if 
all eigenvalues corresponding to the correlation operators 
$\{K^G_a, a\in A_j\}$ of the group $A_j$ (determined by 
Eq.~(\ref{eq:measurement})) are 0, or to discard it otherwise. 
\end{itemize}
Let us take a close look at the sub-protocol ${\mathcal P}_j$.
In (1-1), the multilateral CNOT operation, 
\be \label{eq:create Gj}
\prod_{\{a_1,a_2\}\in A_j} \operatorname{CNOT}^{(a_2\rightarrow a_1)} 
\prod_{\{b_1,b_2\}\in A_{\bar{j}}} \operatorname{CNOT}^{(b_1\rightarrow b_2)},
\ee
is applied between a first state, corresponding to the graph $G$, and a second state corresponding to the graph $g_j$ (from now on, we use $a_l$ for the vertex of the $l$-th state 
in the group $A_j$ and $b_l$ for the one in the group $A_{\bar{j}}$).
A straightforward calculation using Eq.~(\ref{eq:CNOT}) gives the following map
\be
|{\bm \mu_j, \bm\mu_{\bar j}}\rangle_G|{\bm \nu_j, \bm \nu_{\bar j}}\rangle_{g_j} \rightarrow |{\bm \mu_j,\bm \mu_{\bar j}\oplus \bm \nu_{\bar j}}\rangle_G|{\bm \nu_j \oplus \bm\mu_j ,\bm \nu_{\bar j}}\rangle_{g_j},
\ee
where $\oplus$ means bitwise addition modulo 2, which shows the transfer of information about the stabilizer eigenvalues between the two states. 
Note that the final states correspond to the same graphs as the input states.
After that, one measures the second state in order to determine the indices  ${\bm \nu}_j \oplus {\bm \mu}_j$ corresponding to color $A_j$, using the procedure described in (1-2), where
each bit $\nu_j \oplus \mu_j$ is determined using Eq.~(\ref{eq:measurement}).
If all the parities ${\bm \nu}_j \oplus {\bm \mu}_j$  are $\bm{0}$, it is 
expected that $\mu_a = 0$ and $\nu_a = 0$ for $a \in A_j$ are probable since 
$\ket{\bm 0}_{G}$ and $\ket{\bm 0}_{g_j}$ have been assumed to be the majority 
in their ensembles.
That is why then the first state is kept and otherwise discarded. 
As consequence, in the expansion (\ref{mixedgraph}) of the density matrix, elements of the form $\lambda_{\bm{0},\bm{\gamma}_{\bar{j}}}$ are increased. One finds that the new matrix elements of $\rho'_G$ are given by
\begin{align}
{\lambda}'_{\bm{\gamma}_j,\bm{\gamma}_{\bar{j}}}=\frac{1}{\kappa}\sum_{\left\{(\bm{\mu}_{\bar{j}},\bm{\nu}_{\bar{j}})\mid \bm{\mu}_{\bar{j}}\oplus \bm{\nu}_{\bar{j}}=\bm{\gamma}_{\bar{j}}\right\}} \lambda_{\bm{\gamma}_j,\bm{\mu}_{\bar{j}}}\tilde{\lambda}_{\bm{\gamma}_j,\bm{\nu}_{\bar{j}}}
\end{align}
where $\kappa$ is a normalization constant guaranteeing that 
$\operatorname{tr}(\rho')=1$, and $\tilde{\lambda}$ are the coefficients for the two-colorable graph-state $\rho_{g_j}$ written in the form (\ref{mixedgraph}). We remark that here we do not address the question of the unfavorable scaling behavior of the efficiency of the proposed protocol with the total number of particles $N$, for recent developments on this subject in the context of two-colorable graph state purification see \cite{Go06}.

We illustrate how the protocol works by looking at the simple toy-case where noise acts only on one color. Let us consider $G$ as the 5-qubit ring and $g_1$ as the 5-qubit cluster, and study the effect of $\mathcal{P}_1$ on mixed states of the form
\be
\rho_G=\sum_{\bm{\mu}_1}\lambda_{\bm{\mu}_1,\bm{0},0}\,\ket{\bm{\mu}_1,\bm{0},0}_G\bra{\bm{\mu}_1,\bm{0},0},
\ee
\be
\rho_{g_1}=\sum_{\bm{\nu}_1}\tilde{\lambda}_{\bm{\nu}_1,\bm{0},0}\,\ket{\bm{\nu}_1,\bm{0},0}_{g_1}\bra{\bm{\nu}_1,\bm{0},0}.
\ee
Note that even though $g_1$ is two-colorable, we group the vertices in three distinct sets corresponding to the colors of the ring. As only the qubits in $A_1$ are noisy, a sequence of applications of sub-protocol $\mathcal{P}_1$ is sufficient to purify the state. A straightforward calculation gives the new coefficient of the purified ring-state: $\lambda'_{\bm{\gamma}_1,\bm{0},0}=\lambda_{\bm{\gamma_1},\bm{0},0}\,\tilde{\lambda}_{\bm{\gamma_1},\bm{0},0}/\sum_{\bm{\nu}_1}( \lambda_{\bm{\nu_1},\bm{0},0}\,\tilde{\lambda}_{\bm{\nu_1},\bm{0},0})$, meaning that the dominant coefficients are increased. 

The whole purification protocol consists of a sequence of applications of the sub-protocols ${\mathcal P}_j$ corresponding to all colors $j=1,\dots,k$. Even though there is a back-action of noise for the colors which are not purified for the step $j$, one obtains an overall increase of the fidelity $\lambda_{\bm{0}}$ if the fidelity of the initial state is sufficiently high. In fact, $\lambda_{\bm{0}}=1$ is an attractive fix point of the protocol under the ideal local operations.

\subsubsection{Preparation of the auxiliary state $\rho_{g_j}$ in static scenario} \label{sec: preparation of the auxiliary states}

Next we describe how to obtain the auxiliary two-colorable graph-states $\rho_{g_j}$ required for sub-protocol ${\mathcal P}_j$. In the communication scenario, these states can be generated directly by distributing them through (noisy) quantum channels. In the static scenario, the situation is slightly more complicated and a pre-processing to the purification sub-protocol ${\mathcal P}_j$ is needed which we will describe in the following.

Assume that an ensemble of noisy $k$-colorable graph states $\rho_G$ is 
available. We supply an ensemble of the auxiliary two-colorable 
graph state $\rho_{g_j}$ by LOCC:

\begin{itemize}
\item[(0-1)] The parties take two identical noisy copies of $\rho_G$.
The parties in the group $A_j$ apply a CNOT gate from the second (control) copy 
to the first (target) copy, the parties in the group 
$A_{\bar{j}}$ apply the CNOT in the opposite direction.

\item[(0-2)] The parties measure the second state in the eigenbasis of $\sigma_z$. By this, they erase all the edges between the two states and they are left with a state corresponding to the two-colorable graph $g_j$ after a suitable change of local basis depending on the measurement outcome. 
\end{itemize}
A straightforward calculation using Eq.~(\ref{eq:CNOT}) shows that the multilateral CNOT applied in step (0-1) results in the following map when applied to a product of two $k$-colorable graph-states
\be \label{eq:map create Gj}
\ket{\bm{\mu}_j,\bm{\mu}_{\bar{j}}}_G\ket{\bm{\nu}_j,\bm{\nu}_{\bar{j}}}_G\hspace{4cm}\nonumber\\
\mapsto \!\!\! \prod_{\substack{b_1 \in A_{\bar{j}} \\ b'_2 \in N_{b_2}\cap A_{\bar{j}}}}\!\!\!\!\!\!\! \Lambda Z^{(b_1 b'_2)} \ket{\bm{\mu}_j,\bm{\mu}_{\bar{j}}\oplus \bm{\nu}_{\bar{j}}}_{g_j}\ket{\bm{\nu}_j\oplus \bm{\mu}_j,\bm{\nu}_{\bar{j}}}_G,
\ee
which is a product of a graph-state associated to $g_j$ (state 1) and a graph-state associated to $G$ (state 2), with additional edges between the two graphs. The local $\sigma_z$ measurements on the second state (step (0-2)) erase the corresponding vertices and all edges associated to them in the graph. This ensures that after the measurement the first state is a two-colorable graph state corresponding to the graph $g_j$ as desired. 

Note that the multilateral CNOT does not only create the desired two-colorable graph-state, it also results in a transfer of information between the two states. In particular, the part of the index bit of state 2 corresponding to color $A_j$ is given by $\bm{\nu}_j\oplus\bm{\mu}_j$, where $\bm{\mu}$ and $\bm{\nu}$ correspond to the states before the CNOT operation. We are therefore not only able to create the two-colorable graph-state but also to perform at the same time a first step of purification. To this aim, we replace the measurement in the eigenbasis of $\sigma_z$ by a measurement of the correlation operators $K_a^{G}$ with $a\in A_j$, where we keep the state only if the expectation values of all these correlation operators are zero. That is, step (0-2) can be replaced by
\begin{itemize}
\item[(0-2)']
The parties measure the second state locally in the eigenbasis of $\sigma_x$ for the group $A_j$ and in the eigenbasis of $\sigma_z$ for the group $A_{\bar{j}}$.
The first state is kept if all eigenvalues corresponding to correlation operators 
$\{K^G_a, a\in A_j\}$ of the group $A_j$ (determined by 
Eq.~(\ref{eq:measurement})) are 0, and discarded otherwise.
\end{itemize}
Note that depending on the measurement outcomes a local unitary operation on the remaining copy is required to ensure that $\ket{\bm 0}_{g_i}$ is the dominant component of the resulting state $\rho_{g_j}$. To be precise, the parties should apply a local unitary operation 
$\prod_{b_1 \in A_{\bar{j}},\;\;  b'_2 \in N_{b_2}\cap A_{\bar{j}}} 
(\sigma^{(b_1)}_z)^{\zeta_{b'_2}}$ depending on their measurement pattern
$\zeta_{b'_2}$ after either (0-2) or (0-2)'.

%-------------------------------------------------------------------------------------------------
\subsection{Breeding and hashing protocols} \label{sec:hashing}
%-------------------------------------------------------------------------------------------------
Hashing and breeding protocols were introduced for the bipartite Bell state 
in Ref.~\cite{Be96,BDSW96}, for the GHZ state in Ref.~\cite{MaSm02}, and 
generalized to all two-colorable graph states in Ref.~\cite{DAB03,ADB05,Ch04,Ho05}. 
We will now show that the multilateral \textsc{CNOT} operation together with the use of different states allow one to construct hashing and breeding protocols for any graph state. In both cases, we are given an ensemble of $M$ imperfect $k$-colorable graph-states, with $M \rightarrow\infty$. One then transfers information from randomly chosen subsets of $m$ states to perfect states (breeding) or imperfect states (hashing), which are then measured. At each round our knowledge about the remaining states is increased.

We propose here a generalization of the breeding protocol to all graph-states.
A hashing protocol can be obtained in a similar manner with the additional requirement of taking into account the back-action due to the imperfection of the states used to read out the information.
One is given $M$ copies of an $N$-qubit graph-state corresponding to graph $G$, where $G$ is $k$-colorable. We consider $M \rightarrow\infty$. In addition, one is also given $k$ ensembles $E_j$, $j=1,\dots,k$, which one needs to give back at the end, where ensemble $E_j$ contains perfect copies of the two-colorable graph-state corresponding to graph $g_j$. Let us call $B_i$, $i \in {1,\dots,N}$ the binary vector which contains the value of bit $i$ for a sequence of $m$ states. It is possible to determine the parity of all $B_i$ belonging to a given colour $A_j$ simultaneously by measuring only one state. This is done by performing \textsc{CNOT} gates from one state of $E_j$ to the $m$ $k$-colorable graph states for all qubits in $A_j$ and in the opposite direction for the other qubits. To recover the parity of $B_i$, one measures the two-colorable graph-state using the method described in Seq.~\ref{Seq:operations on graph states}. By this, one determines the eigenvalues of all correlation operators corresponding to set $A_j$, which are given by (\ref{eq:measurement}). 
One needs to repeat the procedure at most $M\,S(a_i^{(0)},a_i^{(1)})$ times to obtain perfect knowledge of $B_i$, where $S(a_i^{(0)},a_i^{(1)})=-a_i^{(0)}\operatorname{log}_2a_i^{(0)}-a_i^{(1)}\operatorname{log}_2a_i^{(1)}$ is the entropy and $a_i^{(\mu_i)}=\sum_{\mu_k\neq \mu_i}\lambda_{\mu_1\mu_2\dots\mu_i\dots\mu_N}$. 

\begin{figure}
\includegraphics[width=0.5\textwidth]{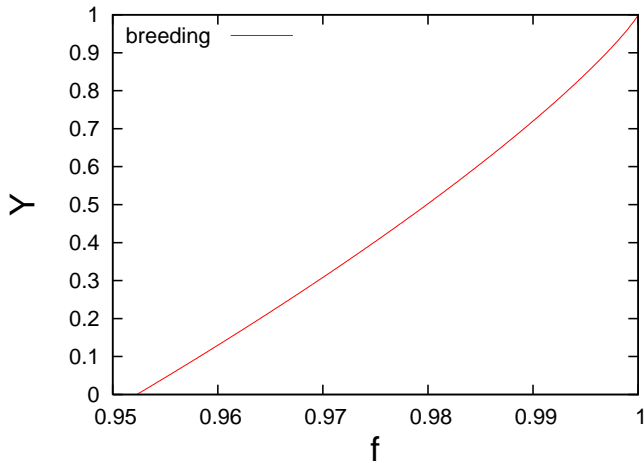}
\caption{Yield as function of fidelity for the breeding protocol applied to the 5-qubit ring-state.}
\label{fig:hashing for N=5}
\end{figure}

For each of the $k$ colors, one performs the sequence of operations described above $\operatorname{max}_{i\in A_j} (M\,S(a_i^{(0)},a_i^{(1)}))$ times in order to obtain the parity of all $B_i$ belonging to this color. Given this information, one ends up with a pure state corresponding (up to local unitary operations) to $M'$ copies of $\ket{\bm{0}}_G$. The last step consists in reconstituting the pool of perfect two-colorable graph-states we were given at the beginning. This is done by performing the multilateral \textsc{CNOT} given by (\ref{eq:create Gj}) on two copies, using by this $2\,\operatorname{max}_{i\in A_j} (M\,S(a_i^{(0)},a_i^{(1)}))$ $k$-colorable graph-states for each color. A lower bound $Y$ of the yield is therefore given by
\be
Y=1-2\,\sum_{j=1}^k\,\operatorname{max}_{i\in A_j} S(a_i).
\ee
We remark that this construction might not be optimal, and a significant improvement of the yield could be achieved by an optimal procedure that generates $M'$ copies $\ket{\Psi_{g_j}}$ from $M$ copies of $\ket{\Psi_{G}}$.
To illustrate our protocol, we calculated the yield for a 5-qubit ring state of the form
\be \label{eq:global white noise on ring} \rho=f\,\ket{\bm{0}}_G\bra{\bm{0}}+(1-f)/(2^5-1)\left(\openone-\ket{\bm{0}}_G\bra{\bm{0}}\right).
\ee
Note that this state is a 3-colorable graph-state.
The entropy $S(a_i^{(0)},a_i^{(1)})$ is identical for all bits. It is given by
\be
S(a_1^{(0)},a_1^{(1)})=\hspace{6cm}&\nonumber\\
-\left(f+(2^4-1)\,\frac{1-f}{2^5-1}\right)\operatorname{log}_2\left(f+(2^4-1)\,\frac{1-f}{2^5-1}\right)\;\;&\nonumber\\
-\left(2^4\,\frac{1-f}{2^5-1}\right)\,\operatorname{log}_2\left(2^4\,\frac{1-f}{2^5-1}\right)\hspace{2cm}&
\ee
The yield, given by $(1-6\,S(a_1^{(0)},a_1^{(1)}))$ is plotted as function of $f$ in Fig.~\ref{fig:hashing for N=5}. One sees that the yield is approaching 1 for a state of fidelity close to 1. 

%-------------------------------------------------------------------------------------------------
\section{Alternative purification protocols via cutting into and reconnecting 2-colorable graph states} \label{sec:alternative}
%-------------------------------------------------------------------------------------------------
Until now we have considered strategies where the $k$-colorable graph-state was directly purified. Another possibility is to purify smaller parts of the graph-state corresponding to two-colorable sub-graphs and connect them at the end. The advantage of this strategy is to make possible the usage of known two-colorable graph state purification protocols. Different intermediate strategies can be designed depending on the chosen sub-graphs. We use an iterative method to construct a set of sub-graphs with no overlapping edges, such that their union is the graph $G$. Step $j$ of the procedure consists of choosing color $A_j$ out of the $k-j$ remaining colors and constructing sub-graph $\tilde{g_j}$, such that $\tilde{g_j}$ is a further sub-graph $g_j$ of Sec.~\ref{graph states:basics} where all vertices $a\in A_{i},i=1,\dots,j-1$ have been erased. At each step, one checks if the graph $G\setminus \{\cup_{i=1}^{j}\tilde{g_i}\}$ is two-colorable. If this is the case, one stops the procedure. The states corresponding to the sub-graphs are distributed from the beginning in a communication scenario, while they are generated from two copies using the multilateral \textsc{CNOT} operation described in Sec.~\ref{sec:recurrence protocol} in a static scenario, followed by a measurement in the $\sigma_z$ basis of the vertices one wants to erase. Once the states are distributed, the multipartite purification protocol for two-colorable graph-states \cite{DAB03,ADB05} is applied. To conclude the protocol, one merges the different graph states together in order to create the final $k$-colorable graph-state. Given two vertices $a_1$ and $a_2$ belonging to different graphs, the corresponding party merges them by applying a projective measurement given by $P_0 = \ket{0}\bra{00}+\ket{1}\bra{11}$ and $P_1 = \ket{0}\bra{01}+\ket{1}\bra{10}$ (with outcomes 0 and 1). A local correction given by $\prod_{b\in N_{a_2}}\sigma_z^{(b)}$ is applied to the state when the measurement result is 1.

\begin{figure}
\setlength{\unitlength}{0.05625\columnwidth}
\begin{picture}(16,8)
\thinlines
\put(0,1.5){\includegraphics[width=0.9\columnwidth]{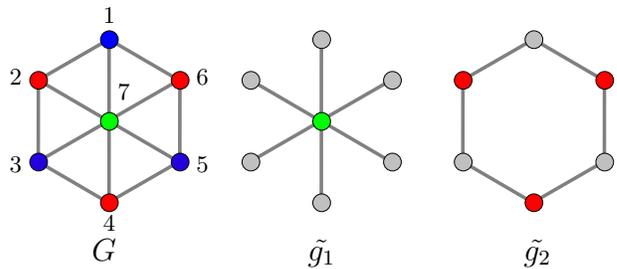}}
\put(2,6.7){1}
\put(-0.55,5){2}
\put(-0.55,2.6){3}
\put(2,1.){4}
\put(4.55,2.6){5}
\put(4.55,5){6}
\put(2.4,4.6){7}
\put(1.7,0.2){\mbox{\large $G$}}
\put(7.6,0.2){\mbox{\large $\tilde{g_1}$}}
\put(13.5,0.2){\mbox{\large $\tilde{g_2}$}}
\end{picture}
\caption{(Color online). A 3-colorable graph $G$ with colors $A_1=\{7\}$, $A_2=\{2,4,6\}$ and $A_3=\{1,3,5\}$. $\tilde{g_1}$ and $\tilde{g_2}$ are two-colorable sub-graph which can be obtained from $G$ and which give $G$ when merging them together. We use the method described in Sec.~\ref{sec:alternative} to construct the sub graphs. 
For example, to obtain $\tilde{g_1}$, we choose $A_1$ as first color and erase all edges between $A_2$ and $A_3$ using the procedure described in Sec.~\ref{sec: preparation of the auxiliary states}. The sub-graph $\tilde{g_2}$ is obtained by erasing the vertex $A_1$ via a $\sigma_{z}$ measurement, and no further processing is required since this graph is already 2-colorable. 
}
\label{fig:sub-graphs construction}
\end{figure}

%-------------------------------------------------------------------------------------------------
\section{Comparison with bipartite strategies}
%-------------------------------------------------------------------------------------------------
\subsection{Performance under ideal local operations in static scenario} \label{sec: comparison with BEPP, perfect LO}
\subsubsection{Yield}
We show in this section that in a static scenario, direct multipartite purification is more efficient than strategies based on bipartite purification. This is due to the fact that any strategy based on bipartite entanglement purification requires that at a certain stage Bell-pairs shared among pairs of parties are generated. After the entanglement purification protocol, another sequence of LOCC must be applied to recover the multiparty entangled states. We show here that these two sequences of operations necessarily generate losses even when applied to pure states. We illustrate this by considering as example the 5-qubit ring state to which we apply the method introduced in Ref.~\cite{ADB05} to quantify the loss. We start with an ensemble of $M$ perfect ring states $\ket{\bm{0}}_G\bra{\bm{0}}$, which are then transformed to an ensemble of Bell pairs $\ket{\bm{0}}^{(kl)}_{\,\,G_2}\bra{\bm{0}}$, shared between the different parties, by means of LOCC. Another sequence of LOCC brings the pairs to an ensemble of $\tilde{M}$ ring states. The total procedure can be summarized as follows:
\be \label{eq:ring->pairs->ring}
\ket{\bm{0}}_G\bra{\bm{0}}^{\otimes M} \rightarrow \bigotimes_{(k,l);k<l}\ket{\bm{0}}^{(kl)}_{\,\,G_2}\bra{\bm{0}}^{\otimes m_{kl}}\rightarrow \ket{\bm{0}}_G\bra{\bm{0}}^{\otimes \tilde{M}}.
\ee
We now calculate a bound on the yield $\tilde{M}/M$ for this procedure. To do it, we apply the following inequalities which were used in Ref.~\cite{LPSW99} to show the irreversibility of entanglement transformation between singlets and GHZ states: (a) The entropy can only decrease on average under LOCC operations; (b) The average increase of relative entropy of system $B=V\setminus A$ is smaller than the average decrease of entropy in system $A$. If we consider a density operator $\rho_{AB}$ describing a pure state which is transformed to an ensemble $\{p_k,\tilde{\rho}^{(k)}_{AB}\}$ by LOCC operations we have that (a)
\be \label{eq:S(rhoA)}
S(\rho^A)\geq \sum_{k} p_k\,S(\tilde{\rho}^A_k),   
\ee
where $S(\rho^A)$ is the von Neumann entropy of the reduced density operator for system $A$, $\rho$ is the initial state and $\tilde{\rho}$ the final state. We consider the bipartition of the system of qubits into system $A$ and system $B=V\setminus A$. Then, we have (b)
\be \label{eq:inequality Er}
\sum_{k}p_k E_r(\tilde{\rho}^{B}_k)-E_r(\rho^{B})\leq S(\rho^A)-\sum_{k}p_k S(\tilde{\rho}^A_k),
\ee
where $E_r(\rho^A)$ is the relative entropy of entanglement for $\rho^A$
\be
E_r(\rho^A)=\min_{\sigma^A {\rm separable}} S(\rho^A||\sigma^A),
\ee
with
\be
S(\rho^A||\sigma^A):=\operatorname{Tr}(\rho^A\operatorname{log}\rho^A)-\operatorname{Tr}(\rho^A\operatorname{log}\sigma^A)
\ee
being the relative entropy. 
Let us now apply this inequalities to our example. We first calculate (\ref{eq:S(rhoA)}) for the second part of process (\ref{eq:ring->pairs->ring}) with $A=\{a_1,a_2\}$ and $B=V\setminus A$. We have for the state describing the ensemble of pairs $S(\rho^A)=\sum_{b;b\in B} \left(m_{a_1b}+m_{a_2b}\right)$. This comes from the fact that the entropy of a pure state is zero and that $\operatorname{Tr}_{B}\rho$ is not a pure state if and only if a single of the 2 qubits belonging to set $A$ is part of an entangled pair. In this case, the entropy is equal to unity for a single state. As there are $m_{ab}$ states containing a fully entangled pair between qubits $a$ and $b$, the contribution of this states to the total entropy is $m_{ab}$. For the $\tilde M$ ring states we find $S(\tilde{\rho}_{A})=2\tilde{M}$, as $\operatorname{Tr}_{B}(|{\bm 0}\rangle_G\langle {\bm 0}|)$ can be written as a sum of 4 projectors with equal weights (see below), and hence for a single copy of the ring state, the entropy of entanglement with respect to the bipartition in question is two. Summing up the contributions of all bipartitions of 2 and 3 qubits, we obtain 
\be
6\,\sum_{a,b;a<b}m_{ab}\geq 20\,\tilde{M}.
\ee
We now apply inequality (\ref{eq:inequality Er}) to the first part of process (\ref{eq:ring->pairs->ring}), with $A=\{a_1,a_2\}$ and $B=V\setminus A$, to have a bound  on $M$. We distinguish two kinds of bipartitions: (i) qubits $a_1$ and $a_2$ are neighbors (ii) they are not neighbors. Let us study the entanglement properties of the state $|{\bm 0}\rangle_G\langle {\bm 0}|$ obtained by tracing out $A$. In case (i), $\operatorname{Tr}_{A}(|{\bm 0}\rangle_G\langle {\bm 0}|)=1/4\,(\ket{000}_{G'}\bra{000}+\ket{001}_{G'}\bra{001}+\ket{100}_{G'}\bra{100}+\ket{101}_{G'}\bra{101})$, where $G'$ stands for the graph corresponding to the 3-qubit GHZ state. A straightforward calculation shows that this state is separable as it can be written as $\operatorname{Tr}_{A}(|{\bm 0}\rangle_G\langle {\bm 0}|)=(\operatorname{Had}\otimes \openone \otimes \operatorname{Had})^{\dag}\,\rho'\,(\operatorname{Had}\otimes \openone \otimes \operatorname{Had})$, with $\rho'=1/4\,(\ket{+++}\bra{+++}+\ket{--+}\bra{--+}+\ket{-+-}\bra{-+-}+\ket{+--}\bra{+--})$. Similarly in case (ii) we find $1/4\,(\openone\otimes \operatorname{(\openone+i\,\sigma_z)} \otimes \operatorname{(\openone+i\,\sigma_z)})^{\dag}\, \operatorname{Tr}_{A}(|{\bm 0}\rangle_G\langle {\bm 0}|)(\openone\otimes \operatorname{(\openone+i\,\sigma_z)} \otimes \operatorname{(\openone+i\,\sigma_z)})=(\openone\otimes \operatorname{Had} \otimes \operatorname{Had})^{\dag}\,\rho'\,(\openone\otimes \operatorname{Had} \otimes \operatorname{Had})$. Hence the state obtained after tracing out 2 qubits is always separable, which implies that the relative entropy $E_r$ vanishes. In addition, we have $S(\rho_A) = S(\rho_B)$ and using the decomposition above we obtain $S(\operatorname{Tr}_{B}(|{\bm 0}\rangle_G\langle {\bm 0}|))=2$.
The relative entropy of entanglement of the ensemble of pairs is given by $E_r(\tilde{\rho}^{B})=\sum_{\{b,c\}\in B}m_{bc}$ as $E_r=1$ for a fully entangled pair and $E_r=0$ for a separable state. We sum up the contributions of the $\binom{5}{2}$ possible bipartitions to obtain 
\be
9\,\sum_{a,b;a<b}m_{ab} \leq 20\,M.
\ee 
Joining the two inequalities we get
\be
\tilde{M}\leq\frac{2}{3}M.
\ee
and hence the procedure (\ref{eq:ring->pairs->ring}) generates losses. On the other hand, breeding presented in Sec.~\ref{sec:hashing} (or equivalently hashing) gives yield 1 for states of fidelity 1, and for a state of form (\ref{eq:global white noise on ring}) with $f>0.9877$ we have $Y > 2/3$. Hence, any 5-qubit ring-state arising from the application of global depolarizing noise and with fidelity $f>0.9877$, can be purified more efficiently using the direct protocol.

\subsubsection{Minimal required fidelity}

In this section, we illustrate the process consisting in generating one entangled pair from a 5-qubit ring-state. We see that the minimal required fidelity is higher for the bipartite strategy. We start with a state resulting from the application of global white noise to the 5-qubit ring-state, given by Eq.~\ref{eq:global white noise on ring}. We define $x=f-\frac{1-f}{2^5-1}$ and rewrite the state as
\be
\label{rhox}
\rho=x\,\ket{\bm{0}}_G\bra{\bm{0}}+\frac{(1-x)}{2^5}\,\openone,
\ee
where $G$ stands for the ring. To create a 2-qubit entangled pair from the initial state, we measure three consecutive qubits in the eigenbasis of $\sigma_z$. We remark that this is in fact an optimal strategy for states of the form Eq. (\ref{rhox}) when operating on a single copy. The resulting state of the remaining two qubits, after some local correction if the measurement result is 1, is given by
\be
\rho=x\,\ket{\bm{0}}_{G_2}\bra{\bm{0}}+\frac{(1-x)}{2^2}\,\openone,
\ee
where $G_2$ is the graph consisting in two vertices connected by an edges.
This state is equivalent up to local unitary operations to a Werner state with fidelity $F=(3x+1)/4$. For $x>1/3$ the state is distillable since bipartite entanglement purification protocols can be successfully applied. The state has positive partial transpose for $x\leq 1/3$ which implies that it is not distillable. We thus have the condition $x>1/3$ such that the state is distillable via a bipartite entanglement purification strategy. For the direct multiparty entanglement purification protocol proposed in this article, we (numerically) find a threshold $x>0.2$ for states of the form Eq. (\ref{rhox}). Thus the minimum required fidelity for the multiparty strategy is significantly lower than for the bipartite strategy.  
Although here we illustrate the advantage of our method by the 5-qubit ring state, such advantages are expected for other graph states as well.

\subsection{Performance under imperfect local operations} \label{sec: comparison with BEPP, imperfect LO}
After having shown the advantage of multipartite purification with respect to bipartite purification in the static scenario, we turn to a more general setting including noisy local operations.
We show here that when local operations are imperfect, multipartite purification can be advantageous also in the communication scenario.

We model noise in the communication channels and in the local operations
as follows. 
We study typical noise models, where the Kraus representation of the superoperators is diagonal in the Pauli basis. This is a common and usually sufficiently general model \cite{DHCB05} (in particular, any noisy channel can be brought to such a form by means of (probabilistic) local operations). We consider the depolarizing channel:
\begin{multline} 
\mathcal{E}^{(a)}_p(\rho)=p\rho+\frac{1-p}{4}\left(\rho+\sigma_x^{(a)}\rho\sigma_x^{(a)}+\right.\\
\left. \sigma_y^{(a)}\rho\sigma_y^{(a)}+\sigma_z^{(a)}\rho\sigma_z^{(a)}\right) \label{eq:depolarizing channel}
\end{multline}
where $p$ is the \textit{reliability}. 
As part of the purification protocols, local one- and two-qubit unitary operations are employed which may be noisy. An imperfect operation is modeled by preceding the perfect operation $U^{(a_1 a_2)}$ with the application of the noise superoperators $\mathcal{E}$ from Eq. (\ref{eq:depolarizing channel}) with parameter $p_l$, i.~e. the state is transformed as
\begin{equation} 
\rho \mapsto U^{(a_1 a_2)} \left( \mathcal{E}^{(a_1)}_{p_l}\mathcal{E}^{(a_2)}_{p_l} (\rho)\right) U^{\dagger(a_1 a_2)}. 
\end{equation} 
We assume that the protocols are executed with the least possible number of operations to keep accumulated noise low. Hence, if a local two-qubit gate $U_{12}^{(a_1 a_2)}$ is preceded by one-qubit gates $U_1^{(a_1)}$ and $U_2^{(a_2)}$ we apply one combined unitary $U^{(a_1 a_2)}=U_1^{(a_1)}U_2^{(a_2)}U_{12}^{(a_1 a_2)}$ which is subjected to noise only once.

We compare a strategy using direct multiparty entanglement purification (MEPP strategy) with a strategy using bipartite entanglement purification (BEPP strategy) in the particular example of the 5-qubit ring-state, which is genuinely three-colorable and hence can be purified directly only by means of the new protocol. To compare BEPP and MEPP strategies, we computed the maximal reachable fidelity $F_{max}$, and the minimum required fidelity $F_{min}$. $F_{max}$ is the maximum value of fidelity to which a state of fidelity $F>F_{min}$ can be brought using the purification protocol. Note that $F_{max}$ is the same in a communication and in a static scenario. This comes from the fact that the maximal reachable fidelity does not depend on the initial state. That is, if the initial state is distillable, it is always possible to go to a state of fidelity $F_{max}$. 

\begin{figure}[t]
\includegraphics[width=0.5\textwidth]{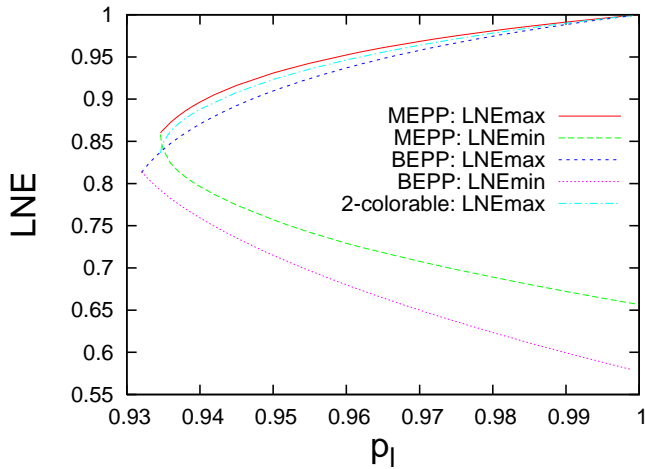}
\caption{Local Noise Equivalent (LNE) corresponding to the maximal reachable fidelity and to the minimal required fidelity as function of the amount of local noise $p_l$ for the 5-qubit ring state in the communication scenario. Given a state of fidelity $F$, the LNE is the level of local depolarizing noise which has to be applied to the perfect state to obtain fidelity $F$. The red solid line and the green dashed line stand for the LNE of the  maximal reachable fidelity and the minimal required fidelity respectively for the MEPP strategy. The same values are plotted for the BEPP strategy. The blue small dashed line is the LNE of a ring state teleported using 4 purified pairs and the pink dotted line 
is th LNE corresponding to the minimal required fidelity of a pair obtained by sending 
one of its qubits through a depolarizing channel. 
Not that the maximal reachable fidelity is the same in a communication and a static scenario. We also plotted the maximal reachable fidelity for the alternative strategy where a 5 qubit cluster-state and a Bell pair are purified before being connected. This value is given by the light-blue dotted-dashed line.}
\label{fig:Fmin Fmax for 5 qubit ring communication}
\end{figure}

\subsubsection{Communication scenario}

Let us describe the MEPP and the BEPP strategies in the communication scenario. The initial states are different in both strategies which renders the comparison non-obvious. We use the local noise equivalent (LNE), which is, for a state of fidelity $F$, the level of local depolarizing noise defined as in Eq.~(\ref{eq:depolarizing channel}), which has to be applied to the perfect state to obtain fidelity $F$.  

In the MEPP strategy, the setting is the following: 5 parties A,B,C,D, and E are connected by depolarizing channels; the party A creates states locally and distributes them to the four other parties. To use direct purification, in addition to the 5-qubit ring states, the party A needs to distribute 5-qubit cluster-states in three different ways, allowing the purification with respect to the three different colors. The purification is done in two steps. The parties first purify the three different cluster states up to their maximum reachable fidelity, before using them to purify the ring-states up to $F_{max}^{MEPP}(p_l)$, where $p_l$ gives the amount of local noise. The fact that $F_{max}^{MEPP}(p_l)$ is always reached for a state with $F>F_{min}^{MEPP}(p_l)$ is then used to compute $F_{min}^{MEPP}(p_l)$. For a given amount of local noise $p_l$, we vary the channel noise $q$ to find the threshold value $q_{min}$ above which the state can be purified up to $F_{max}^{MEPP}(p_l)$. $F_{min}^{MEPP}(p_l)$ is obtained by applying the depolarizing channel with noise parameter $q_{min}$ to 4 of the qubits of the ring-state.

In the BEPP strategy, the party A creates Bell pairs and distributes them to the other parties. The pairs are purified and then used to teleport locally created 5-qubit ring-states. We adopted a conservative scenario where one of the 5 parties creates the ring-states, decreasing by this the number of Bell pairs needed to teleport the states from 5 to 4. In addition, we assume that the teleportation process itself does not add additional imperfections/noise. Hence the actual value of $F_{max}^{BEPP}$ is lower than our conservative estimate.
To obtain the maximal reachable fidelity of multiparty entangled states $F^{BEPP}_{max}(p_l)$, we purify the Bell pairs up to  $F^{Bell}_{max}(p_l)$ and use $N-1$ of them to teleport a locally created ring-state. 
The results presented in Fig.~\ref{fig:Fmin Fmax for 5 qubit ring communication}, show that in a communication scenario, the minimum required fidelity as well as the threshold value $p_l$, under which no purification is possible is always lower in the BEPP strategy. However, the maximal reachable fidelity is higher in the MEPP scenario allowing us to get states of higher fidelity. 

We also studied the alternative strategy consisting in distributing two-colorable sub-graph states which are purified and connected at the end. There are two different possibilities depending on the choice of the first color in the construction of the sub-graphs. If one chooses an ensemble of two qubits as first color, one gets a 5-qubit cluster-state and a 2-qubit cluster state (5-2 strategy), while if the first color contains only 1 qubit, one gets a 3-qubits cluster-states plus a 4-qubit one (4-3 strategy). The classification of the different strategies by decreasing value of the maximal reachable fidelity is as follows: direct purification, 5-2 strategy, 4-3 strategy and finally the BEPP strategy. The order is inverted for the value of minimal required fidelity.

\begin{figure}[t]
\includegraphics[width=0.5\textwidth]{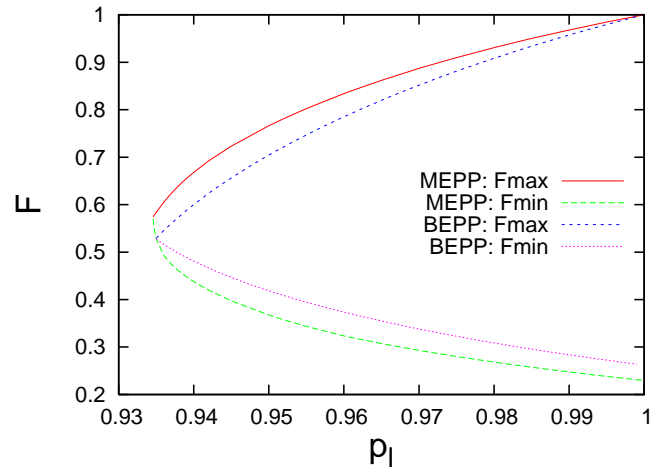}
\caption{Maximal reachable fidelity and minimal required fidelity as function of the amount of local noise for the 5-qubit ring state in the static scenario. The red solid line and the green dashed line stand for the maximal reachable fidelity and the minimal required fidelity respectively, for the MEPP strategy, while the blue small dashed line and the pink dotted line stand for the same quantities in the BEPP scenario. The value of $F_{min}$ plotted for the BEPP scenario is the minimum value of fidelity of the ring state, so that the pairs we extract from the ensemble of rings can be purified.}
\label{fig:Fmin Fmax for 5 qubit ring static}
\end{figure}

\subsubsection{Static scenario}

A similar calculation can be done for the static scenario. In this case, one is given an ensemble of distributed 5-qubit ring-states described by $\rho=\prod_{a=1}^{4}\mathcal{E}^{(a)}_p\ket{\bm{0}}_G\bra{\bm{0}}$. As noted in the previous subsection, the maximal reachable fidelity is the same as in the communication scenario. In the MEPP strategy, the minimum required fidelity is similar in both scenarios. Indeed, the creation of the 3 different cluster states can be done together with a first purification with respect to one of the colors. The behaviour of the protocol in the static scenario therefore doesn't differ much from the one in the communication scenario if one begins the purification of the two-colorable graph-state corresponding to graph $g_j$, by sub-protocol $\mathcal{P}_j$. However, the minimum required fidelity is different in the BEPP strategy. It is defined as the minimum value of fidelity of the ring-state, such that at least one distillable pair can be extracted from it. The results are presented in Fig.~\ref{fig:Fmin Fmax for 5 qubit ring static}. The MEPP strategy clearly presents an advantage in terms of fidelity. The minimum required fidelity is lower and the maximal reachable fidelity larger for any value of reliability $p_l$.

%-------------------------------------------------------------------------------------------------
\section{Conclusion}
%-------------------------------------------------------------------------------------------------

In this paper, we have proposed multipartite entanglement purification
(recurrence and breeding) protocols by which parties can distill
arbitrary graph state directly.
The work not only gives a complete systematic package for the construction
of entanglement purification protocols for graph states, but also clarifies
a special role of two-colorable graph states in reading out the parity
information of the stabilizer eigenvalues of any graph state.
The latter property might open a new avenue to a ``patch-work'' purification
of decohered qubits in resource entangled states for quantum information
processing. 
We remark that very recently a similar idea, i.e. the usage of different shapes of graphs, has been utilized by Goyal, McCauley and Raussendorf to obtain purification protocols for two--colorable graph states with improved yield and scaling behavior \cite{Go06}.

We have considered two scenarios, namely (i) the static LOCC scenario and (ii)
the communication scenario. Under ideal local operations, we have showed in the
static scenario that our protocol gives a higher yield and a wider distillable
regime (i.e., a smaller required fidelity for distillability), compared with
any bipartite strategy.
Also, under noisy local operations, our protocol allows one to distill
mixed states up to a higher achievable fidelity in both scenarios.

%%%%%%%%%%%%%%%%%%%%%%%%%%%%%%%%%%%%%%%%%%%%%%%%%%%%%%%%%%%%%%%%%%%%%%%%%%%%%%
\section*{Acknowledgements}
We thank Simon Anders and Caterina Mora for helpful discussions.
This work was supported by the FWF, the European Union (OLAQUI, SCALA, QICS), the \"OAW through project APART (W.D.) and JSPS (A.M.).
%%%%%%%%%%%%%%%%%%%%%%%%%%%%%%%%%%%%%%%%%%%%%%%%%%%%%%%%%%%%%%%%%%%%%%%%%%%%%%

\end{document}